# Non-Ergodicity & Microscopic Symmetry Breaking of the Conductance Fluctuations in Disordered Mesoscopic Graphene

G. Bohra[1], R. Somphonsane[2], N. Aoki[3], Y. Ochiai[3], R. Akis[4], D. K. Ferry[3,4], and J. P. Bird[1,3]

1: Department of Electrical Engineering, University at Buffalo, the State University of New York, Buffalo, NY 14260-1920, USA

2: Department of Physics, University at Buffalo, the State University of New York, Buffalo, NY 14260-1500, USA

3: Graduate School of Advanced Integration Science, Chiba University, 1-33 Yayoi-cho, Inage-ku, Chiba 263-8522, Japan

4: School of Electrical, Computer, and Energy Engineering, Arizona State University, Tempe, AZ 85287-5706, USA

We show a dramatic deviation from ergodicity for the conductance fluctuations in graphene. In marked contrast to the ergodicity of dirty metals, fluctuations generated by varying magnetic field are shown to be much smaller than those obtained when sweeping Fermi energy. They also exhibit a strongly anisotropic response to the symmetry-breaking effects of a magnetic field, when applied perpendicular or parallel to the graphene plane. These results reveal a complex picture of quantum interference in graphene, whose description appears more challenging than for conventional mesoscopic systems.

The understanding of transport in condensed matter has been challenged by the isolation of graphene, whose energy bands have a linear dispersion and whose carriers possess intrinsic isospin [1,2]. These features give rise to a variety of unusual transport phenomena [3-8], not observed in typical semiconductors. Graphene also exhibits pronounced mesoscopic [9-13] effects – including weak localization (WL) [14-19], conductance fluctuations (CF) [20-38], and quantum noise [39-43] – with much richer characteristics than their counterparts in non-Dirac materials [9-13]. Most notably, while the CF in dirty metals and semiconductors are known to exhibit specific universal properties, recent studies [24,25,30,35] suggest that this universality may not extend to graphene. It is this specific issue that we focus on here.

The universal conductance fluctuations (UCF) exhibited by dirty metals [10-13] arise when carrier coherence is preserved on a scale comparable to the system size, in which limit the conductance does not ensemble average but fluctuates deterministically with magnetic field and/or Fermi energy. Vital to understanding these fluctuations, which are much larger than would be expected classically, is the introduction of a so-called *ergodic hypothesis* [11,12]. This proposes that CF observed while sweeping magnetic field or Fermi energy should be equivalent to those obtained by varying the disorder potential. Based on this hypothesis, the CF are expected to have universal amplitude at zero-temperature, independent of sample size and the degree of disorder. At non-zero temperatures, where the characteristic lengths describing coherent transport can become smaller than the sample size, the UCF are damped, albeit in a manner well described by theory [13].

Recent theories for the CF in graphene [44-48] predict universal behavior in the metallic regime ($k_F l \gg 1$, where $k_F$ is the Fermi wavevector and $l$ the mean free path), after accounting for inter-valley scattering arising from different sources of disorder [45,46]. The universality is inferred, however, from an implicit assumption of ergodicity, something that recent experiment has called into doubt [35]. Given the importance of the ergodic hypothesis for our understanding of mesoscopic transport, it is essential to establish its applicability to graphene. In this Letter, we therefore perform a detailed study of the CF in disordered graphene, while varying Fermi energy (carrier density) and magnetic field. We find that these parameters generate CF with different amplitudes, implying a failure of the ergodic hypothesis. We furthermore use a magnetic field to investigate the response of the CF to the breaking of microscopic symme-tries [49-54], revealing a strong, and unexpected, anisotropy, dependent on the orientation of this field relative to the graphene sheet.

Devices were fabricated by exfoliating [2] natural graphite onto a Si/SiO$_2$ substrate. The SiO$_2$ was 300-nm thick, and the heavily-doped Si served as a back-gate that could sweep carrier density/Fermi energy. Substrate markers ensured accurate alignment for electron-beam lithography, allowing Cr/Au (5-/40-nm) contacts to be made to graphene flakes. These were selected by optical and atomic-force microscopy, and by Raman spectroscopy. As indicated in Fig. 1(a), our measurements utilized a two-probe configuration most appropriate for studies of UCF [13]. *I-V* curves between these contacts were linear over a wide range [56], suggesting contact resistance provided only a small contribution to the measurements [57]. After wire-bonding, samples were mounted on the cold finger of a dilution refrigerator, either perpendicular ($B_\perp$) or parallel ($B_\parallel$) to the magnetic field. Measurements of the Hall effect in GaAs/AlGaAs mesas have shown that this allows alignment of the sample with respect to the external field to an error of no more than 0.2°. Results were first obtained in the perpendicular configuration, following which we warmed to 300 K and switched to the in-plane orientation. CF were measured by low-frequency lock-in detection, with a constant current of 0.7 nA, and, unless stated otherwise, the cryostat temperature was 0.04 K.

Among six different devices studied, four used monolayer graphene while a further two were bilayer. The Dirac point of these devices was systematically shifted to positive voltage (arrows in Figs. 1(b) & 1 (c)), indicative of chemical doping. 4.2-K conductivity at the Dirac point ranged from ~1.5 – 4.0 × $e^2/h$, and mobility far away from this point was around 400 – 1000 cm$^2$/Vs. For the same conditions we find that $k_Fl \approx 1.5$, indicating that the samples are strongly disordered and far from the metallic limit ($k_Fl \gg 1$) considered elsewhere [45,46]. (For more details on the characterization see the supplementary material.) This strongly-disordered character likely reflects the fact that neither thermal annealing, nor current-induced cleaning, were performed to remove any organic residues left over after fabrication. Prior work [57] has demonstrated the presence of such residues in uncleaned devices, and has shown them to introduce significant disorder into the graphene transport. We therefore emphasize that the interesting behaviors reported here should likely be considered to be strongly related to the dirty nature of the graphene.

Figures 1(b) & 1(c) show, respectively, the gate-voltage dependent resistance ($R$) of a bilayer (length $L$ = 2 µm, width $W$ = 4 µm) and monolayer ($L$ = 4 µm, width $W$ ≤ 1 µm) device. As reported previously [20-38], CF grow significantly as the temperature is lowered towards 4.2 K. Their reproducibility is highlighted in Fig. 1(d), which shows their dependence on temperature in the monolayer device (in the remainder of this paper, we focus on results obtained for this device). The CF ($\delta g$, in units of $e^2/h$) were obtained by subtracting a slowly-varying background from the raw conductance. Even at the lowest temperature, the CF are small ($\delta g < e^2/h$), which we attribute to the characteristic length scales governing mesoscopic transport [13]. From the diffusion constant ($D \simeq 50$ cm$^2$/s, supplementary material), we infer a thermal diffusion length of $L_T = \sqrt{\hbar D / k_B T} \geq 1$ µm at 0.04 K. Similarly, by calculating the correlation field [13] of the magneto-CF at 0.04 K, we estimate a phase-breaking length ($l_\varphi$) of around 200 nm. Thus, our experiment is performed in a regime where $l_\varphi \ll L_T$, $W$ & $L$, where decoherence should significantly suppress the CF.

In Fig. 2(a) we compare, on the same scale, CF from two different experiments. The upper curves show magneto-conductance ($-2 \leq B_\perp \leq +2$ T) at several gate voltages, while the lower curves show conductance versus back-gate voltage (i.e. the transconductance) at different magnetic fields ($1.0 \leq B_\perp \leq 1.8$ T). It is apparent to even the naked eye that Fig. 2(a) *shows strongly non-ergodic CF*. That is, CF generated by sweeping Fermi energy (back-gate voltage) are clearly larger than those obtained by sweeping magnetic field. The non-ergodicity is *not* some artifact of the manner in which we chose to select the different data sets. Magneto-conductance was measured for the gate voltages indicated by the colored arrows at the bottom-left corner of Fig. 2(a), over which range the transconductance shows a large variation. The small CF observed in the magnetic-field sweeps therefore do not correspond to some fortuitous range of back-gate voltage, for which the transconductance fluctuations are also small. Moreover, we emphasize that while the curves of Fig. 2(a) have been shifted vertically for clarity, they otherwise correspond to *as-measured* data. Consequently, differences in the two types of fluctuations cannot be attributed to some error in background subtraction.

In Fig. 2(b), we plot the root-mean square conductance fluctuation ($\delta g_{rms}$, in units of $e^2/h$), determined from the magneto-conductance (between ±2.0 T) at different gate voltages (red data, lower axis), and from the transconductance (measured between 0- & 60-V) at different $B_\perp$ & $B_\parallel$ (blue data, upper axis). Comparing with the trans-CF at zero field, $\delta g_{rms}$ obtained from the magneto-conductance is around a factor of three smaller, independent of the gate voltage. Even if we compare the field-induced fluctuations with those obtained by varying Fermi energy at $B_\perp$ = 2.0 T, the difference in amplitude is around a factor of two. This is still significantly larger than any error bars in the data, confirming the non-ergodicity suggested in Fig. 2(a).

Another key feature of the UCF is their response to the breaking of microscopic symmetries. A magnetic field may be used to break time-reversal and/or spin degeneracy, with implications that have been demonstrated for mesoscopic metals and semiconductors [49-54]. In Fig. 3, we present measurements of the transconductance fluctuations in the graphene device as a function of $B_\perp$ & $B_\parallel$. Figs. 3(a) & 3(c) plot these results as color contours, whose common scales allow a direct comparison of the influence of the field direction on the CF. In Figs. 3(b) & 3(d), in contrast, we show resistance as a function of gate voltage at three representative magnetic fields. Although the fluctuation fingerprints differ markedly between these figures, this can be attributed to the fact that the device was thermally cycled to room temperature between the measurements. The contour of Fig. 3(a) shows a strong influence of $B_\perp$ on the CF, which are clearly largest over the narrow range -0.5 < $B_\perp$ < +0.5 T, where the most pronounced color variation is obtained. $\delta g$ is suppressed outside of this range, as indicated by the lack of color contrast in this section of the contour. This suppression can also be seen in the line curves of Fig. 3(b), which show a clear decrease of the CF on increasing $B_\perp$ to 6 T (note the indicated offsets in Figs. 3(b) & 3(d)). Very different behavior is apparent in the contour of Fig. 3(c), which shows that the application of in-plane fields up to 6 T has remarkably little influence on the CF. This is confirmed by the line plots of Fig. 3(d), which show only a weak dependence of the CF on $B_\parallel$.

In Fig. 2(b), we plot (as blue data points) the variation of $\delta g_{rms}$ from transconductance measurements in the two field configurations. At zero magnetic field, $\delta g_{rms}$ is essentially equivalent for the two sets of data, providing confidence that the statistical characteristics of the CF are not significantly affected by thermal cycling. With an out-of-plane field applied, we first observe a rapid decrease of $\delta g_{rms}$ by a factor of $1/\sqrt{2}$, when $B_\perp$ is increased to around 0.5 T. In the standard theory of UCF [49-51], this drop is precisely that expected for breaking time-reversal symmetry, consistent with which we note that the field scale on which it occurs correlates well to the width of the WL peak in the magneto-resistance (which we show, for example, at 4.2 K in Fig. 1(e)). Following the rapid drop, $\delta g_{rms}$ then shows a much slower decrease as $B_\perp$ is further increased, before saturating beyond 4 T at a value close to half of its zero-field value. This second decrease in $\delta g_{rms}$, by an additional factor of $1/\sqrt{2}$, is also well known from the study of UCF, and is precisely that expected for lifting spin degeneracy in a system with weak spin-orbit coupling [51,53]. (This nice agreement with theoretical expectations provides further evidence that the analysis of our data is not significantly influenced by contact-resistance, since the observed reduction factors of $1/\sqrt{2}$ and $1/2$ indicate the measured resistance is dominated by that of the graphene flake).

The weak dependence of the CF on in-plane magnetic field is puzzling. $\delta g_{rms}$ appears independent of $B_\parallel$, indicative of only a weak Zeeman splitting for in-plane magnetic fields. Although one possibility is that the *g*-factor is much smaller in the graphene plane, studies of graphite have shown this parameter to be isotropic [58]. A more likely explanation is an extrinsic, substrate-induced, anisotropy, and we note that a recent study of the spin states of graphene quantum dots also found a much weaker spin splitting for $B_\parallel$ [59]. In fact, for $B_\perp$ these authors observed the onset of Zeeman splitting beyond 2 – 4 T, consistent with our data in Fig. 2(b). Anisotropic spin relaxation has furthermore been reported in spin-valve studies [60], where it was attributed to different effective spin-orbit fields for in-plane and out-of-plane relaxation. In contrast to these results, however, clear Zeeman splitting of the CF was observed [29] for an in-plane field in experiments performed on less-disordered samples than those studied here. These different observations point collectively to a strong dependence of the in-plane spin splitting on sample quality.

While early theory [44] questioned the universality of the CF in graphene, later studies predicted them to be universal, albeit with an amplitude that is sensitive to the sources of inter-valley scattering [45,46]. This conclusion was reached, however, by applying usual perturbative treatments [13] to compute CF amplitudes by ensemble averaging. To connect to the results of experiment an ergodic hypothesis is then required, in which the ensemble average is *assumed* to be equivalent to varying magnetic field or Fermi energy. That is, such approaches *inevitably* predict *equivalent* CF in magnetic field or Fermi en-ergy. Our experiments suggest this assumption is not always valid, and that the CF can be non-universal. Such a failure of universality represents a dramatic breakdown in our understanding of mesoscopic phenomena, and presumably requires a more proper treatment of microscopic disorder. While the usual theory of UCF applies when $k_Fl \gg 1$ [13,45,46], we work in the regime where $k_Fl \approx 1$. Here, it is not clear that perturbative approaches [13] to quantum transport remain valid, since conduction is likely influenced by carrier puddling [61-65]. Indeed, a breakdown of ergodicity was suggested in Ref. 35, in which CF generated by magnetic field and gate voltage showed different amplitudes near the Dirac point, where such puddling should be most important. (This difference was less pronounced than that found here, however, where the difference in amplitudes is as much as a factor of eight, see Fig. 2) In our case we observe the failure of ergodicity over the entire range of gate voltage studied, and not just near the Dirac point, which tends to suggest that the influence of the puddling can extend over a wide range of density. One possibility, that could be explored in the future through detailed studies of the differential conductance, is that some incipient Coulomb blockade governs the gate-voltage induced CF, causing them to exhibit a different amplitude to the field-induced features.

Finally, we comment on the short phase-breaking length ($l_\varphi < L_T$) in our samples. A correlation analysis (see supplementary material) suggests that this is primarily the result of a saturation of $l_\varphi$ that onsets around 1 K. A similar saturation has also been reported in experiments on dirty metals and semiconductors [66], and even today its origins are subject to debate. Saturation has furthermore been found in other studies of the CF in graphene [18,19,21,32], with phase-breaking lengths in close agreement with ours. The behavior that we observe therefore appears to be a manifestation of a general phenomenon, whose origins are still not well understood.

In conclusion, mesoscopic interference in disordered graphene exhibits a dramatic breakdown of the ergodic hypothesis, with CF obtained by varying magnetic field being significantly smaller than those obtained when sweeping Fermi energy. These results reveal a complex picture of quantum interference in graphene, whose description appears more challenging than for conventional mesoscopic systems.

This research was supported by the U.S. Department of Energy, Office of Basic Energy Sciences, Division of Materials Sciences and Engineering under Award DE-FG03-01ER45920 (GB & JPB), and by the National Science Foundation under Award OISE0968405 (RS). We also gratefully acknowledge the assistance of J. M. Velazquez and B. J. Schultz with the micro-Raman measurements.

# REFERENCES

1. H. Castro Neto, F. Guinea, N. M. R. Peres, K. S. Novoselov, and A. K. Geim, Rev. Mod. Phys. **81**, 109 (2009).

2. K. S. Novoselov, A. K. Geim, S. V. Morozov, D. Jiang, Y. Zhang, S. V. Dubonos, I. V. Grigorieva, and A. A. Firsov, Science **306**, 666 (2004).

3. K. S. Novoselov, A. K. Geim, S. V. Morozov, D. Jiang, M. I. Katsnelson, I. V. Grigorieva, S. V. Dubonos, and A. A. Firsov, Nature **438**, 197 (2005).

4. Y. Zhang, J. P. Small, M. E. S. Amori, and P. Kim, Phys. Rev. Lett. **94**, 176803 (2005).

5. Y. Zhang, Y.-W. Tan, H. L. Stormer, and P. Kim, Nature **438**, 201 (2005).

6. M. I. Katsnelson, K. S. Novoselov, and A. K. Geim, Nat. Phys. **2**, 620 (2006).

7. A. F. Young and P. Kim, Nat. Phys. **5**, 222 (2009).

8. N. Stander, B. Huard, and D. Goldhaber-Gordon, Phys. Rev. Lett. **102**, 026807 (2009).

9. G. Bergmann, Phys. Rep. **107**, 1 (1984).

10. S. Washburn and R. A. Webb, Adv. Phys. **35**, 375 (1986).

11. B. L. Altshuler, JETP Lett. **41**, 648-651 (1985).

12. P. A. Lee and A. D. Stone, Phys. Rev. Lett. **55**, 1622 (1985).

13. P. A. Lee, A. D. Stone, and H. Fukuyama, Phys. Rev. B **35**, 1039 (1987).

14. S. V. Morozov, K. S. Novoselov, M. I. Katsnelson, F. Schedin, L. A. Ponomarenko, D. Jiang, and A. K. Geim, Phys. Rev. Lett. **97**, 016801 (2006).

15. E. McCann, K. Kechedzhi, V. I. Fal'ko, H. Suzuura, T. Ando, and B. L. Altshuler, Phys. Rev. Lett. **97**, 146805 (2006).

16. X. Wu, X. Li, Z. Song, C. Berger, and W. A. de Heer, Phys. Rev. Lett. **98**, 136801 (2007).

17. R. V. Gorbachev, F. V. Tikhonenko, A. S. Mayorov, D. W. Horsell, and A. K. Savchenko, Phys. Rev. Lett. **98**, 176805 (2007).

18. F. V. Tikhonenko, D. W. Horsell, R. V. Gorbachev, and A. K. Savchenko, Phys. Rev. Lett. **100**, 056802 (2008).

19. D. K. Ki, D. Jeong, J. H. Choi, H. J. Lee, and K. S. Park, Phys. Rev. B **78**, 125409 (2008).

20. C. Berger, Z. Song, X. Li, X. Wu, N. Brown, C. Naud, D. Mayou, T. Li, J. Hass, A. N. Marchenkov, E. H. Conrad, P. N. First, and W. A. de Heer, Science **312**, 1191 (2006).

21. D. Graf, F. Molitor, T. Ihn, and K. Ensslin, Phys. Rev. B **75**, 245429 (2007).

22. S. V. Morozov, K. S. Novoselov, M. I. Katsnelson, F. Schedin, D. C. Elias, J. A. Jaszczak, and A. K. Geim, Phys. Rev. Lett. **100**, 016602 (2008).

23. H. B. Heersche, P. Jarillo-Herrero, J. B. Oostinga, L. M. K. Vandersypen, and A. F. Morpurgo, Nature **446**, 56 (2007).

24. N. E. Staley, C. Puls, and Y. Liu, Phys. Rev. B **77**, 155429 (2008).

25. S. Russo, J. B. Oostinga, D. Wehenkel, H. B. Heersche, S. S. Sobhani, L. M. K. Vandersypen, and A. F. Morpurgo, Phys. Rev. B **77**, 085413 (2008).

26. V. Skákalová, A. B. Kaiser, J. S. Yoo, D. Obergfell, and S. Roth, Phys. Rev. B **80**, 153404 (2009).

27. W. Horsell, A. K. Savchenkoa, F. V. Tikhonenkoa, K. Kechedzhib, I. V. Lernerc, V. I. Fal'ko, Sol. St. Comm. **149**, 1041 (2009).

28. Y Ujiie, S. Motooka, T. Morimoto, N. Aoki, D. K. Ferry, J. P. Bird, and Y. Ochiai, J. Phys.: Cond. Matt. **21** 382202 (2009).

29. M. B. Lundeberg and J. A. Folk, Nat. Phys. **5**, 894 (2009).

30. S. Branchaud, A. Kam, P. Zawadzki, F. M. Peeters, and A. S. Sachrajda, Phys. Rev. B **81**, 121406 (2010).

31. J. Velasco, G. Liu, L. Jing, P. Kratz, H. Zhang, W. Bao, M. Bockrath, and C. N. Lau, Phys. Rev. B **81**, 121407 (2010).

32. Y. Oh, J. Eom, H. C. Koo, and S. H. Han, Sol. St. Comm. **150**, 1987 (2010).

33. K. K. Mahelona, A. B. Kaiser, and V. Skákalová, Phys. Stat. Sol. (b) **247**, 2983 (2010).

34. Y.-F. Chen, M.-H. Bae, C. Chialvo, T. Dirks, A. Bezryadin, and N. Mason, J. Phys.: Condens. Matt. **22**, 205301 (2010).

35. C. Ojeda-Aristizabal, M. Monteverde, R. Weil, M. Ferrier, S. Gueron, and H. Bouchiat, Phys. Rev. Lett. **104**, 186802 (2010).

36. J. Moser, H. Tao, S. Roche, F. Alzina, C. M. Sotomayor Torres, and A. Bachtold, Phys. Rev. B **81**, 205445 (2010).

37. J. Berezovsky, M. F. Borunda, E. J. Heller, and R. M. Westervelt, Nanotechnol. **21**, 274013 (2010); M. F. Borunda, J. Berezovsky, R. M. Westervelt, and E. J. Heller, ACS NANO **5**, 3622 (2011).

38. J. Berezovsky and R. M. Westervelt, Nanotechnol. **21**, 274014 (2010).

39. R. Danneau, F. Wu, M. F. Craciun, S. Russo, M.Y. Tomi, J. Salmilehto, A. F. Morpurgo, and P. J. Hakonen, Phys. Rev. Lett. **100**, 196802 (2008).

40. L. Dicarlo, J. R. Williams, Y. Zhang, D. T. McClure, and C. M. Marcus, Phys. Rev. Lett. **100**, 156801 (2008).

41. R. Danneau, R. Danneau, F. Wu, M.F. Craciun, S. Russo, M.Y. Tomi, J. Salmilehto, A.F. Morpurgo, and P.J. Hakonen, Sol. St. Comm. **149**, 1050 (2009).

42. A. N. Pal and A. Ghosh, Phys. Rev. Lett. **102**, 126805 (2009).

43. G. Xu, C. M. Torres, Jr., Y. Zhang, F. Liu, E. B. Song, M. Wang, Y. Zhou, C. Zeng, and K. L. Wang, Nano Lett. **10**, 3312 (2010).

44. A. Rycerz, J. Tworzydlo, C. W. J. Beenakker, Europhys. Lett. **79**, 57003 (2007).

45. K. Kechedzhi, O. Kashuba, V. I. Falko, Phys. Rev. B **77**, 193403 (2008).

46. M .Yu. Kharitonov and K. B. Efetov, Phys. Rev. B **78**, 033404 (2008).

47. K. Kechedzhi, D. W. Horsell, F. V. Tikhonenko, A. K. Savchenko, R. V. Gorbachev, I. V. Lerner, and V. I. Fal'ko, Phys. Rev. Lett. **102**, 066801 (2009).

48. E. Rossi, J. H. Bardarson, M. S. Fuhrer, and S. Das Sarma, arXiv:1110.5652v1.

49. Y. Imry, Europhys. Lett. **1**, 249 (1986).

50. B. L. Al'tshuler and B. I. Shklovskii, Zh. Eksp. Teor. Fiz. **91**, 220 (1986) [Sov. Phys. JETP **64**, 127 (1986)].

51. K. A. Muttalib, J.-L. Pichard, and A. D. Stone, Phys. Rev. Lett. **59**, 2475 (1987).

52. N. O. Birge, B. Golding, and W. H. Haemmerle, Phys. Rev. Lett. **62**, 195 (1989).

53. P. Debray, J. L. Pichard, J. Vicente, and P. N. Tung, Phys. Rev. Lett. **63**, 2264 (1989).

54. O. Millo, S. J. Kleppner, M. W. Keller, D. E. Prober, S. Xiong and A. D. Stone, Phys. Rev. Lett. **65**, 1494 (1990).

55. See Supplemental Material at [ … ] for an overview of the samples and their basic electrical characterization.

56. X. Du, I. Skachko, F. Duerr, A. Luican, and E. Y. Andrei, Nature **462**, 192 (2009).

57. J. Moser, A. Verdaguer, D. Jiménez, A. Barreiro, and A. Bachtold, Appl. Phys. Lett. **92**, 123507 (2008); J. Moser, A. Barreiro, and A. Bachtold, Appl. Phys. Lett. **91**, 163513 (2008);

58. J. M. Schneider, N. A. Goncharuk, P. Vašek, P. Svoboda, Z. Výborný, L. Smrčka, M. Orlita, M. Potemski, and D. K. Maude, Phys. Rev. B **81**, 195204 (2010).

59. J. Guttinger, T. Frey, C. Stampfer, T. Ihn, and K. Ensslin, Phys. Rev. Lett. **105**, 116801 (2010).

60. N. Tombros, S. Tanabe, A. Veligura, C. Jozsa, M. Popinciuc, H. T. Jonkman, and B. J. van Wees, Phys. Rev. Lett. **101**, 046601 (2008).

61. S. Adam, E. H. Hwang, V. M. Galitski, and S. Das Sarma, Proc. Natl. Acad. Sci. **104**, 18392 (2007).

62. J.-H. Chen, C. Jang, S. Adam, M. S. Fuhrer, E. D. Williams, and M. Ishigami, Nat. Phys. **4**, 377 (2008).

63. J. Martin, N. Akerman, G. Ulbricht, T. Lohmann, J. H. Smet, K. von Klitzing, and A. Yacoby, Nat. Phys.**4**, 144 (2008).

64. S. Jung, G. M. Rutter, N. N. Klimov, D. B. Newell, I. Calizo, A. R. Hight-Walker, N. B. Zhitenev, and J. A. Stroscio, Nat. Phys. **7**, 245 (2011).

65. A. Deshpande, W. Bao, Z. Zhao, C. N. Lau, and B. J. LeRoy, Phys. Rev. B **83**, 155409 (2011).

66. J. J. Lin and J. P. Bird, J. Phys.: Condens. Matter **14**, R501 (2002).

**FIGURE CAPTIONS**

**Fig. 1:** (a) Optical micrograph of the monolayer-graphene device. Measurement probes are indicated and the graphene flake can be seen in the inset. (b), (c) Resistance versus back-gate voltage for the bilayer and monolayer device, respectively, at several temperatures. The arrow in each plot identifies the Dirac point. (d) CF in the monolayer device at different temperatures. (e) Magneto-CF of the monolayer device at several gate voltages (12 – 54 V) at 4.2 K. Each curve has had a smooth background subtracted.

**Fig. 2:** (a) Comparison of CF obtained by sweeping $B_\perp$ (upper curves, upper horizontal axis) and back-gate voltage (lower curves, lower horizontal axis) at a cryostat temperature of 0.04 K. Fixed gate-voltage and magnetic-field values are indicated on the plot. Colored arrows indicate gate voltages at which the magneto-conductance measurements were made. (b) Blue data points show $\delta g_{rms}$ determined from gate-voltage induced CF at different $B_\perp$ (open symbols) and $B_\parallel$ (filled symbols). Red data points show $\delta g_{rms}$ from the magneto-conductance at various gate voltages. To allow the dependence of $\delta g_{rms}$ on $B_\perp$ to be clearly seen, we have slightly displaced the other two data sets by the indicated increments. Dotted lines indicate the range of $B_\perp$ for which $\delta g_{rms}$ drops to $1/\sqrt{2}$ and 1/2 of its zero-field value. Cryostat temperature was 0.04 K.

**Fig. 3:** (a) In this contour, conductance was measured as a function of gate voltage, following which CF were obtained by background subtraction. The evolution of these CF as $B_\perp$ was incremented from –1 to 6 T was then plotted in the contour. (b) Variation of resistance with gate voltage for three representative values of $B_\perp$ (indicated). (c) As in (a) except gate-voltage induced CF are plotted as a function of $B_\parallel$. (d) As in (b) except now for three values of $B_\parallel$. $V_D$ in (a) & (c) denotes the Dirac point. Cryostat temperature was 0.04 K in panels (a) – (d). Curves in (b) & (d) are shifted vertically by the offsets (in kΩ) indicated in the panels.

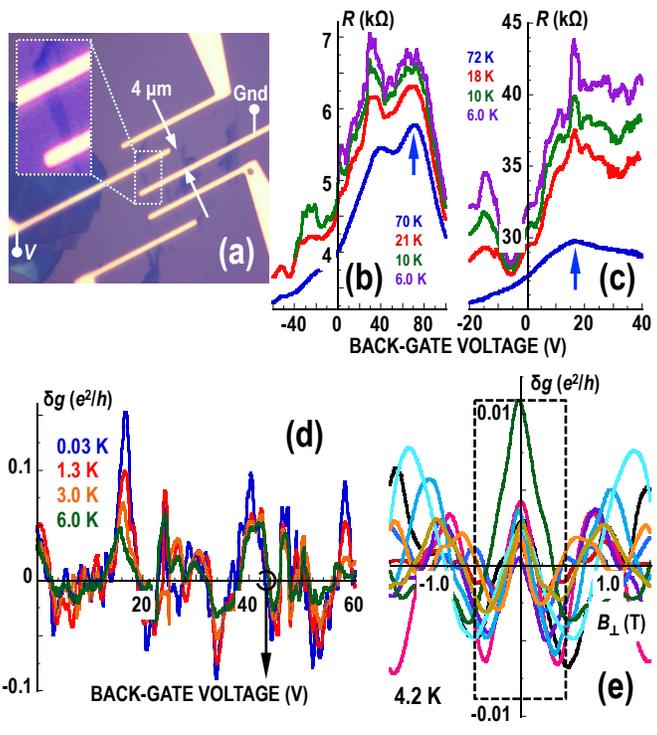

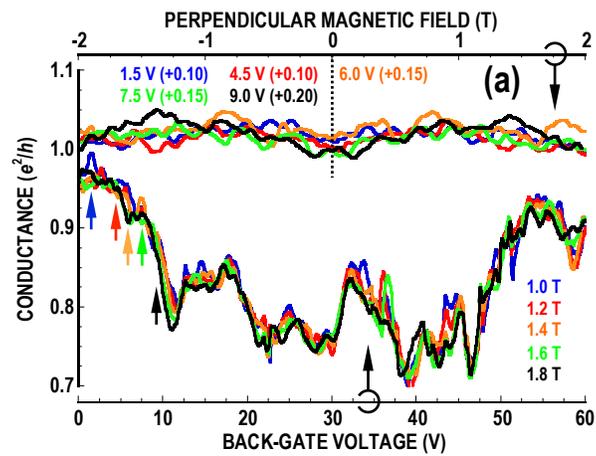

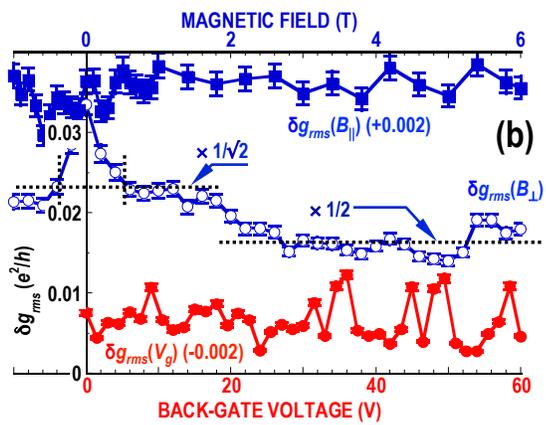

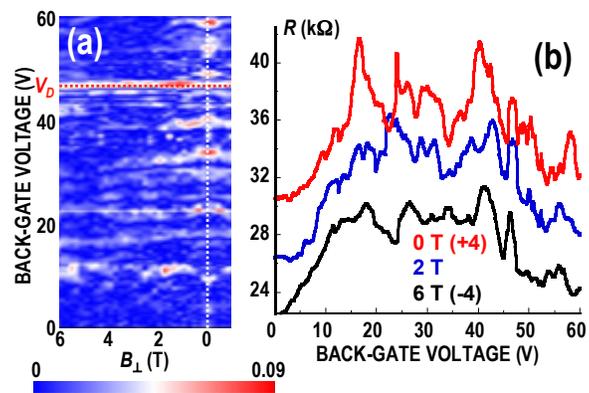
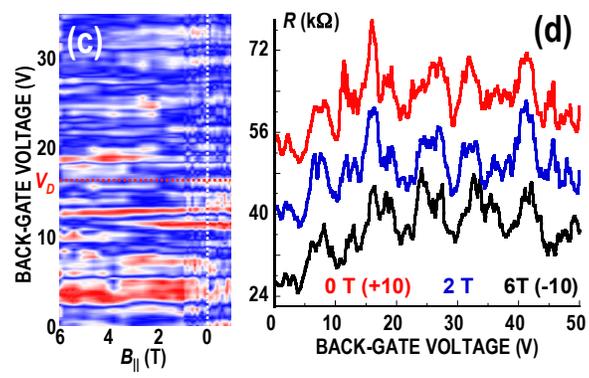